\documentclass[12pt]{article}
\usepackage{graphicx}
\begin{document}

\begin{center}
{\LARGE \bf  Einstein, Kant, and Taoism} \\

\vspace{3ex}

Y. S. Kim \\
Department of Physics, University of Maryland, \\
College Park, Maryland, 20742, U.S.A.

\vspace{3ex}

\end{center}

\begin{abstract}

It is said that Einstein's conceptual base for his theory of
relativity was the philosophy formulated by Immanuel Kant.  Then,
is it possible to see how Kant's philosophy played a role in
Einstein's thinking without reading Kant's books?  This question
arises because it is impossible for physicists to read Kant's
writings.  Yes, it is possible if we use the method of physics.  It
is known also that Kant's mode of thinking was profoundly affected
by the geography of Koenigsberg where he spent eighty years of his
entire life.  We examine what aspect of this geography led Kant to
create his philosophy upon which Einstein's concept of relativity
was based.  It is pointed out that the Eastern philosophy of Taoism
is a product of the geographical environment similar to that
of Kant's Koenigsberg, and therefore that it is easy to absorb
Kantianism with Taoist background.

\end{abstract}

\section{Introduction}

From the philosophical point of view, Einstein was a
Kantianist~\cite{kim95,howard05}.  It is also known that Kant's
philosophy was influenced by the culture of Koenigsberg where he spent
his entire life~\cite{apple94}.  In this article, we examine how
Einstein's thinking was influenced by the lifestyle of Kant's birth
place.

According to Eugene Wigner, philosophers do not dictate others how
to think.  They only interpret how people think~\cite{wig89}.  Then
Immanuel Kant (1724-1804) formulated his philosophy based on the
people he met throughout his life.  Kant was born and spent his entire
life in an East Prussian city called Koenigsberg.  He was strictly
a local person.  Therefore, his line of thinking was thoroughly
configured by the citizens of Koenigsberg during his time.  In 1946,
this city became a Russian city called Kaliningrad.  Therefore,
if Kaliningrad has a feature distinct from other places, we are
interested.

According to Karl Marx, philosophers interpret this world in various
ways.  There comes the question of changing the world~\cite{kmarx}.
Marx was wrong if regarded himself as the philosopher and Vladimir
Lenin as the person who would change the world.  On the other hand,
Marx was right if Kant was the philosopher and Einstein was the person
who changed the world.  Einstein was more than a philosopher.  He was
a physicist.  Thus, philosophers can help us, but we cannot solely
depend on them for providing answers to what Einstein means to us.
We are interested in how Einstein was able to formulate his physical
ideas which changed the world.

In this paper, we are interested in the aspect of Kantianism that
things depend on how observations are made, namely on the observer's
status of mind and on his/her environment.  Modern physics is an
observer-dependent science.  In special relativity, the same physics
can appear differently to observers in different Lorentz frames.
The kinetic energy of a particle in one Lorentz frame can be written
as $E = p^2/2m$ when the particle speed is much smaller than the
speed of light, while it takes the form $E = pc$ when the observer's
frame moves with a speed close to $c.$

In quantum mechanics, we have a wave-particle duality.  For observers
who can see only particles, a particle looks like a particle.  On the
other hand, it looks like a wave for observers who can see only waves.
Quantum mechanics is also an observer-dependent science, within the
framework of Kant's philosophy.

If Kantianism is a product of the geography of Koenigsberg now called
Kaliningrad, the best way to study this geography is to go there and
look at the place and talk to the people who live there and those who
used to live there.  For this purpose, I went there in June of 2005
and spent three days and two nights.  Indeed, based on my observations,
I am now able to assert that Kantianism, at least the aspect which
affected Einstein, was a product of the geography of the city which
served as Kant's entire world.

Again, if we believe in Kantianism, the contents of this article depend
heavily on who the author is, and how his brain was configured by his
environments.  The author of the present paper was born in Korea and
came to the United States in 1954 after high school graduation.  He
received both undergraduate (Carnegie-Mellon) and graduate (Princeton)
degrees in the United States.  In addition, he has been teaching
American students since 1962 after joining the faculty of the
University of Maryland.  He is in a position to compare Kant-oriented
Koreans with American scientists with the Edisonian tradition.

In Sec.~\ref{geo}, I report what I observed and learned while visiting
Kant's city now called Kaliningrad.  In Sec.~\ref{histo}, based on
the observation I made there, we trace the history of Koenigsberg which
led Kant to formulate his philosophy.  We list in Sec.~\ref{illus}
various examples in our daily life which can be explained in terms of
Kantianism.  In Sec.~\ref{phys}, it is noted that modern physics is
an observer-dependent science.  It is then pointed out that the
origin of this observer dependence can be traced to Immanuel Kant who
constructed his philosophy based on the lifestyle of Koenigsberg where
everything is observer-dependent.

In Sec.~\ref{tao}, we trace the history to see that Taoism originated
in ancient China is based on different observers with two opposite
viewpoints.  I then venture to say that its historical origin is the
same of that of Kantianism.  We note in Sec.~\ref{einam} that Einstein
did not communicate well with American physicists.  The reason could
be that Einstein was a Kantianist while the philosophy of America is
yet to be defined.  Let us tentatively call the American philosophy
``Edisonism.''  The distinction between these two philosophies is
spelled out.

I started a systematic research on Einstein's Kantian connection
in 1995 while reading a book review by Marie Arana-Ward in the
Washington Post, an influential newspaper published in the capital
city of the United States.  She was reviewing a book written about
Kaliningrad by Anne Applebaum~\cite{apple94}.  Her review was much
shorter than the book, and is to the point.  With her permission I
include her article in the Appendix of the present paper.

\section{Geography of Kaliningrad}\label{geo}
Immanuel Kant was born in an East Prussian city of Koenigsberg in
1724 and lived there until he died in 1804.  He left the city once
for a brief period to attend his father's funeral.  If Kant's style
of thinking was configured by the society where he grew up and
worked, he had contacts only with the citizens of Koenigsberg.
Then what aspect of Koenigsberg was so distinct from the rest of
the world?

We learn history in order to understand what is happening today.
Likewise, one way to understand history is to study what is happening
now to construct a picture of the past.  With this point in mind, let
us see what is going on in Koenigsberg these days.  Since Kant wrote
his books in German, we are tempted to locate Koenigsberg within the
territory of Germany, but it does not exist there.  Furthermore, the
city is now a Russian city called Kaliningrad, located at the Baltic
wedge between Poland and Lithuania.  Since 1945, Kaliningrad had been
an important submarine base for the Soviet Union until recently.  It
still serves as a naval base for Russia.

The best way to study the geography of a city is to go there and look
at the city and its surrounding areas.  Indeed, for this purpose, I
went to Kaliningrad in June of 2005.  Because, the city's primary
function used to be to serve as a sensitive military base for the
Soviet Union, it was impossible to go there during the Soviet era.
The city is still an isolated place.

Before going to Kaliningrad, I attempted to contact a physics professor
at the University of Kaliningrad, but could not find anyone.  Thus, I
had to arrange my visa and hotel accommodation through the Intourist,
a Russian tourist company which has been operating since Stalin's
Soviet era.  This tells us how isolated this city is.  There are no
railway connections from Poland, and I had to fly from Warsaw on a
propeller-driven airplane, operated by a branch of the LOT Polish
airline.  Its flight schedule was erratic.  It is possible for Russians
there to go to Moscow by overnight trains, but they need transit visas
to Lithuania and Latvia.  There are no civilian airports to accommodate
jet airlines.  In short, it is not yet completely safe to travel to
Kaliningrad.

When I arrived at the airport, a German lady whom I met on the plane
asked me whether I am interested in going to the city with her Russian
friend who came to the airport to pick her up.  I said yes, because
the place was totally strange to me, and I would save sixty dollars
on taxi fare.  Both of them spoke German.  This was precisely what I
was expecting to see.  Kaliningrad is now a Russian city, but its
German root still persists.  I stayed at a hotel called ``Kaliningrad''
located at the center of the city.  Most of the hotel guests came
from Germany.  Many of them were born there before 1945, and I enjoyed
talking with them.  I am also one of those who had to flee in 1946
from my Korean hometown to avoid Soviet troops.

Before 1945, Koenigsberg was an important outpost for Germany,
especially for those German politicians with a big ambition of
expanding eastward.  During World War II, Germans built powerful
radio transmitters there to preach those in Baltic nations how great
Germany is.  While Soviet troops were advancing toward Berlin in 1945,
Hitler ordered all the 400,000 inhabitants of the Koenigsberg region
to move to western provinces of Germany, but only one half of them
were able to flee and the other half were trapped by Stalin's Soviet
army.

Those trapped Germans were all sent to post-war construction sites
in the Soviet Union, and the city became filled with Russians.  The
Lomonosov campus of Moscow State University was built by German
laborers.  With an understanding of Churchill and Truman at the
Potsdam conference (July 1945), Stalin annexed Koenigsberg and
its surrounding areas to the Soviet Union, and changed the name of
the city to Kaliningrad.  It was a big prize for Stalin because it
gave him a much needed non-freezing harbor for his Baltic fleet.

What happened to Immanuel Kant known to and respected by the entire
world?  Germans of course are very proud of Kant, who wrote his books
in the language they speak.  They were not allowed to visit Kant's
grave during the Soviet era.  How about Russians?  Russian couples
usually have their wedding ceremonies on Saturdays.  After the
ceremony, they visit the grave of their hero they respect most.
Kant's grave site becomes very crowded with newlywed couples on
Saturdays.  Indeed, Kant is one of the Russian heros.  How do I know
this?  I was at Kant's grave site and saw them when I was in
Kaliningrad.

Kant's grave is located outside an old church building located at the
city center of Kaliningrad.  Many prominent persons from Koenigsberg
were buried inside the church.  Kant was buried outside because he did
not believe in Jesus.  There is a museum dedicated to Kant in one of
the church turrets.  Russians, even during the Soviet era, did their
best to preserve the historical items relating to Immanuel Kant.  We
shall continue this story of Kant museum in Sec.~\ref{tao}.

After the collapse of the Soviet Union, the Russian government did
not and still does not have enough resources to impose a strong
influence on the region, and foreigners are allowed to visit the city
of Kaliningrad.  Naturally, many Germans come and spend money there.
Russians love German money but want to maintain their firm control
over the city and surrounding region.  On the other hand, Germans
like to have something for their money.  They want to have their city
back.

Germans built a brewery there and produce beer trademarked
``Koenigsberg,'' and I drank a bottle of Koenigsberg while I was in
Kaliningrad.  It was very good.  Until 2005, their university was
called ``Kaliningrad State University,'' but it was renamed to ``Kant
State University.''  It is very safe to assume that Germans contributed
a large sum of money to the university for this name change.  German
capitalists are very busy in constructing resort facilities along
the Baltic coast of the Kaliningrad region.

For the same city, Germans and Russians have two different views.
It is unlikely that there will be a violent armed struggle often seen
in other parts of the world.  Then it would be interesting to see how
these two different views can be blended into the destiny of
Kaliningrad.  One place with the same destiny with two different
ways of looking at the same place!  With this background, we can
trace the history of Kaliningrad or Koenigsberg.

\section{History of Koenigsberg}\label{histo}

In 2005, there was a celebration in Kaliningrad to commemorate the
750-th anniversary of the founding of the city.  President Vladimir
Putin of Russia and Chancellor Gerhardt Schroeder of Germany attended
the ceremony marking this occasion.  Exactly 750 years before, a rich
and strong man built a castle at the top of a mountain there and
proclaimed himself to be the king.  This is the reason why the place
was called king's mountain.  However, the land and its people existed
many years before this Koenig.  What did they do?  What happened to
them?

If we look at the map containing, Kaliningrad, Lithuania, Poland,
Belarus, and Ukraine, there are no natural barriers to isolate
themselves from foreign invaders.  Thus, anyone with a strong army
could march into the area and impose his own management and his own
cultural values.  For this reason, the people there had to develop
skills to accommodate many different ways of looking at themselves.
I learned this from Applebaum's book entitled {\it Between East and
West: Across the Borderlands of Europe}~\cite{apple94}, and its
comprehensive book review by Marie Arana-Ward in the Washington Post
(November 1994).

Then, what makes Kaliningrad different from Poland or Ukraine?  The
key difference is that this area has a large lagoon which can provide
a harbor for the vessels navigating in the Baltic Sea.  Thus, the city
served as a fishing center for centuries.  Unlike Poland or Ukraine,
this coastal area is blessed with generous rains.  This allows grasses
to grow providing good food to cattles.  Kaliningrad was a center for
cattles producing high-quality meats.  While I was there, I enjoyed
a steak of generous size.

Kaliningrad is also blessed with abundance of amber.  There are many
amber mines along the Baltic coast.  Katherine the Great of Russia
built an amber room in her palace from the ambers from this area.
The original amber room was taken away by German troops during World
II when they surrounded the city of Leningrad and occupied Katherine's
palace located at the place now called ``Pushkin.''  It is believed
that the amber room in its disassembled form is buried in Koenigsberg,
but its location is not known.  A remake of this amber room is now
in the Hermitage Museum of Saint Petersburg.

Thus, before politicians started giving names to the place such as
Koenigsberg or Kaliningrad, this harbor city was a commercial center
for the Baltic Sea, as Venice was for the Mediterranean Ocean.  The
region became richer and stronger, while accommodating many different
points of view from different people from different places.  The area
surrounding the city became a small country called Prussia, with
Koenigsberg as its capital city.

Prussia became so rich that it was able to purchase a large chunk of
land west of Poland including the city now known as Berlin.  Since
this new area was close to Western European countries going through
industrial revolution, it developed very rapidly, and Prussia's center
of gravity moved to Berlin.  The original Prussia then became absorbed
into a new empire called Deutschland, and it became a province called
East Prussia.  In Poland, this region is still called Prussia.

While there was a rich flow of history, Koenigsberg had enough
resources to construct and maintain a top-class university, cultivating
many outstanding scholars.  Depending on political climate of the city,
this university went through different names.  The present name is
{\it Kant State University}.  Until 1945, the university constantly
maintained its excellent academic tradition consistent with Immanuel
Kant's creativity.  The university went through a dark age when the
city was closed during the cold-war era.  From 1945 to 2005, the
university was called {\it Kaliningrad State University}.  It served
as a local institution providing collage education to young Russians
from the Kaliningrad region.  These days, both Russians and Germans
seem to agree on the need for reconstruction of its academic tradition.

In physics, the University of Koenigsberg was the birth place of
mathematical physics.  We know how to write down Coulomb's law for two
charged particles.  The same law can be written in terms of electric
field and Gauss's law.  Gauss's law was extended to cover all Maxwell's
equations.  The integral form of Maxwell's equations was completed at
the University of Koenigsberg, as Arnold Sommerfeld states clearly
in his book on electrodynamics~\cite{sommer52}.  Sommerfeld was a
student there.

\vspace{3mm}

\begin{table}[thb]

\caption{Although Minkowski formally declared the Lorentz covariance
of Maxwell's equations 1907, Einstein knew it before 1905 and was
afraid of the physical world in which Newton's mechanics and
electrodynamics have two different transformation properties.  To
those who know only Maxwell's electrodynamics, this world is Lorentzian.
To those who know only Newton's mechanics, the same world is Galilean.
Einstein wanted to have one transformation law for both mechanics and
electrodynamics.}\label{maxein}

\begin{center}
\begin{tabular}{ccc}
{}&{}&{}\\
Space and Time \hspace{3mm}  {} & \hspace{5mm}
Maxwell \hspace{5mm}&\hspace{5mm}
Newton \hspace{5mm}  \\[2mm]\hline
{}&{}&{}\\
Lorentzian \hspace{3mm}  & YES & NO \\[4mm]\hline
{}&{}&{}\\
Galilean \hspace{3mm} & NO  & YES \\[4mm]\hline
\end{tabular}

\end{center}
\end{table}
\vspace{3mm}

While those Koenigsbergers were busy in studying Maxwell's equations,
Hendrik Lorentz of the Netherlands and Henri Poincar\'e of France were
working on the mathematics known today as the Lorentz group.  Hermann
Minkowski from Lithuania was a graduate student at the University of
Koenigsberg at that time.  He became interested in whether Maxwell's
equations and their solutions are consistent with the Lorentzian world.
He continued working on this problem while he was a professor at the
University of Munich, where Einstein was a student.  Einstein was not
a good student to him (he was never a good student to anyone), but
he sensed from Minkowski the Lorentz covariance of electrodynamics,
even though Minkowski did not publish his work in 1907.  Then, how
about Newtonian mechanics?  This is the question which confronted
Einstein, as is illustrated in Table~\ref{maxein}.

As I emphasized in Sec.~\ref{geo}, this is the type of question the
citizens of Koenigsberg had to face throughout their history.  This
is precisely what Kant's philosophy is about.  Indeed, Kantianism is
a product of the lifestyle of Koenigsberg.

\section{Illustrative Examples of Kantianism}\label{illus}
In order to gain a concrete grasp of Kantianism, let us examine some
examples in the real world which can be explained within the framework
of Kantianism.  To physicists, it is impossible to understand physical
or philosophical theories without seeing what happens in the real
world.  We shall discuss here limitation of observations, observations
with a fixed frame of mind, different manifestation of the same object,
and construction of the true world based on different observations.

The Italian city of Trieste is known to physicists as the home of the
International Center for Theoretical Physics.  The center is about
five kilometers from the downtown, and there is a bus connection along
the beach where many beautiful girls swim and relax.  When I was on
the bus with my friend, it was half full, and I could see everybody
including the driver.  I told my friend everybody on this bus in woman.
He then told me it is so because my eyes can detect only women.  My
friend was a physicist and knew how to speak the language of modern
physics.  I never found out whether it was so because there were only
women or because of measurement problems with my eyes.

At a dinner time, I said that Trieste is an excellent place to stage
a war.  It has sea, mountains, islands, and beautiful ladies.  A
perfect combination!  Of course, my comment outraged everybody, but
military professionals would agree with me.  In fact, many generals
in the past chose this place to fight.  Then what do I have in common
with those military people?  The answer is very simple.  I was born
and raised in Korea.  When I was born, Korea was hosting Japanese
troops.  American and Soviet troops after 1945.  Then the Korean War
(1950-53) before I came to the United States 1954 after my high
school graduation.  This means that my body and mind were configured
in the war environment.  When I weigh things, I still compare them
with the weight of the M-1 rifle of the U.S. army weighing ten pounds
and its bayonet weighing one pound.

About 4,500 years ago, there was a king named Yao in China.  While
he was looking for a man who could serve as his prime minister, he
heard from many people that a person named Shiyu was widely respected
and had a deep understanding of the world.  The king then sent his
messengers to invite Shiyu to come to his palace and run the country.
After hearing the king's message, Shiyu, without saying anything,
walked down to a creek in front of his house and started washing his
ears.  He thought he heard the dirtiest story in his life.

Shiyu is still respected in the Eastern world as one of the wisest men
in history.  We do not know whether this person existed or is a made-up
personality.  In either case, we are led to look for a similar person
in the Western world.  In ancient Greece, each city was run by its
city council.  As we experience even these days, people accomplish very
little in committee meetings.  Thus, it is safe to assume that the
city councils in ancient Greece did not handle matters too efficiently.
For this reason, there was a well-respected wise man, like Shiyu, who
never attended his city council meetings.  His name was Idiot.  Idiot
was a wise man, but he never contributed his wisdom to his community.
His fellow citizens labeled him as a totally useless person.  This was
how the word idiot was developed in the Western world.

Idiot and Shiyu had the same personality if they were not the same
person.  However, Idiot is a useless person in state-centered societies
like Sparta.  The same person is regarded as the ultimate wise man in a
self-centered society like Korea.  I cannot say that I know everything
about other Asian countries, but I have a deep knowledge of Korea where
I was born and raised.  The same person looks quite differently to
observers in different cultural frames.  While living in the United
States with my Eastern background, I was frequently forced to find a
common ground for two seemingly different views.

As an example closer to physics, let us look at a Coca-Cola can.  It
appears like a circle if we look at it from the top, while it looks
like a rectangle from its side.  The real thing is a three-dimensional
circular cylinder.

Immanuel Kant emphasized the importance of the observer's subjective
viewpoint in his book entitled ``Kritik der reinen Vernunft'' whose
first and second editions were published in 1781 and 1787 respectively.
While Kant was interested in finding the real thing from different
observations, he became to believe in absoluteness of the real thing.

However, using his own logic, he ended up with a conclusion that there
must be the absolute inertial frame, and that we only see the frames
dictated by our subjectivity.  He noted that there are many moving
bodies in the universe.  To him, the earth also was a moving object,
like all other planets or stars.  He then became obsessed with the
absolute frame of reference.  There are still many who believe in the
absolute frame.  Here we cannot blame Kant too much for failing to
come up with the principle of relativity.  He did not have the benefit
of Lorentz-covariant formulation of Maxwell's equations.

We all know how Einstein's view is different from the Kantian view
of the universe.  However, without Kantian philosophical background,
it could have been very difficult for Einstein to formulate his
relativity.  He introduced a Lorentzian dimension for the time
variable.

\section{Kantian Influence on Modern Physics}\label{phys}

Unlike classical physics, modern physics depends heavily on observer's
state of mind or environment.  The wave-particle duality is a product
of Kantianism.  If your detector can measure only particle properties,
particles behave like particles.  On the other hand, if your detector
can detect only wave properties, particles behave like waves.
Heisenberg had come up with the uncertainty principle to reconcile
these two different interpretations.  This question is still being
debated, and is a lively issue these days.

In special relativity, observers in different Lorentz frames see the
same physical system differently.  The energy-momentum relation for
a massive particle is $E = p^2/2m$ if it is at rest.  However, an
observer moving with a speed close to that of light would insist that
the same relation should be $E = cp$.  It was Einstein who was able
to settle the quarrel between these two observers, as is illustrated
in Table~\ref{wigein}~\cite{kim89}.  This is a manifestation of
Kantianism.

\begin{table}[thb]
\caption{Further contents of Einstein's $E = mc^{2}$.  Massive and
massless particles have different energy-momentum relations.
Einstein's special relativity gives one relation for both.  Wigner's
little group unifies the internal space-time symmetries for massive
and massless particles.  The covariant harmonic oscillators can
explain why the quark model and the parton model are two different
manifestations of the same Lorentz-covariant entity.}\label{wigein}.

\begin{center}
\vspace{3mm}

\begin{tabular}{cccc}

{}&{}&{}&{}\\
{} & Massive, Slow \hspace*{1mm} & COVARIANCE \hspace*{1mm}&
Massless, Fast \\[4mm]\hline
{}&{}&{}&{}\\
Energy- & {}  & Einstein's & {} \\
Momentum & $E = p^{2}/2m$ & $ E = \sqrt{(cp)^{2} + (mc^{2})^{2}}$ & $E = cp$
\\[4mm]\hline
{}&{}&{}&{}\\
Internal & $S_{3}$ & {}  &  $S_{3}$ \\[-1mm]
Space-time &{} & Wigner's  & {} \\ [-1mm]
Symmetry & $S_{1}, S_{2}$ & Little Group & Gauge
Trans. \\[4mm]\hline
{}&{}&{}&{}\\
Relativistic & {} & Covariant Model & {} \\[-1mm]
Extended & Quark Model & of & Partons \\ [-1mm]
Particles & {} & of Hadrons & {} \\[4mm]\hline

\end{tabular}

\end{center}
\end{table}


Einstein, in 1905, formulated his special relativity for a point
particle.  It became known that particles have internal space-time
symmetries.  Massive particles have the spin degree of freedom
when they are at rest.  If they move fast or become massless, they
preserve the spin along the direction of momentum, but the transverse
components become contracted into one gauge degree of freedom.  Is
it possible to combine these into one Lorentz-covariant theory?
This work was started in Wigner's 1939 paper on representations
of the Lorentz group~\cite{wig39}, but was not completed until
1990~\cite{kiwi90jm}.  The second row of Table~\ref{wigein}
illustrates this aspect of Kantianism.

The concept of spin comes from quantum mechanics.  Thus, Wigner's
work constitutes an important component of the task of combining
quantum mechanics with Einstein's special relativity.  The remaining
problem is whether quantum mechanics, with its probability
interpretation, is consistent with Einstein's relativity.  Many
people believe that the present form of quantum field theory is the
answer to this question.  They are partially right.

In order to combine quantum mechanics with relativity, we should know
how to deal with its wave picture in a Lorentz-covariant manner.
It is a trivial matter to build plane waves covariantly starting from
the Klein-Gordon equation.  The present form of quantum field theory
leading to the Lorentz-covariant S-matrix and Feynman diagrams is
based on those plane waves.  However, because of this, field theory
cannot explain standing waves in a Lorentz-covariant manner.

Let us look at the hydrogen atom.  It looks like a localized
probability entity if the atom is on the table.  How would it look
to an observer on a bicycle?  The correct answer to this question is
``We do not know.''  This has been an agonizing question to Paul A. M.
Dirac.  He examined the role of the c-number time energy relation,
light-cone coordinate system, and harmonic oscillators for
representations of the Lorentz group.  If we combine all of these
ideas, it is possible to construct oscillator wave functions which
can be Lorentz-boosted~\cite{knp86}.  It is then possible to work out
the third row of Table~\ref{wigein} by showing that the covariant
oscillator produces three-dimensional harmonic oscillators for the
quark model for slow hadrons and the parton model for fast-moving
hadrons.

\section{Kantianism and Taoism}\label{tao}

During the years 1985 - 1991, I went to Princeton every two weeks to
tell Professor Eugene Wigner stories he wanted to hear.  Of course I
was telling him about physics.  In order to do this, I had to tune
my mode of thinking to Wigner's way of reasoning.  I once asked him
whether he thinks like Immanuel Kant.  He said Yes.  I then asked
him whether Einstein was a Kantianist.  Wigner said ``Yes, Einstein
was definitely a Kantianist.''  I then asked Wigner whether he studied
Kant's philosophy while he was in college.  He said No, and said that
he realized he had been a Kantianist after writing so many papers in
physics.  This conversation took place in 1989~\cite{wig89}.

He added that philosophers do not dictate others how to think, but
their job is to describe systematically how people think.  Here,
Wigner was talking about Immanuel Kant who formulated his philosophy
based on the thinking mode of the inhabitants of Koenigsberg.  Wigner
then told me that I was the only person who asked him this question,
and asked me how I knew the Kantian way of reasoning was working in
his mind.  I gave him the following answer.

I never had any formal education in oriental philosophy, but I know
that my frame of thinking is affected by my Korean background.  One
important aspect is that Immanuel Kant's name is known to every
high-school graduate in Korea, while he is unknown to Americans.
When I mention Kant's name to American physicists, they dismiss him
as a philosopher having nothing to do with physics.  The question
then is whether there is something common between Eastern culture
and Kantian philosophy.

I would like to answer this question in the following way.  Koreans
absorbed a bulk of Chinese culture during the period of the Tang
dynasty (618-907 AD).  At that time, China was the center of the
world as the United States is today.  This dynasty's intellectual
life was based on Taoism which tells us, among others, that
everything in this universe has to be balanced between its plus
(or bright) side and its minus (or dark) side.  This way of thinking
forces us to look at things from two different or opposite directions.
This aspect of Taoism could constitute a ``natural frequency'' which
can be tuned to the Kantian view of the world where things depend
how they are observed.

Later in 1180 AD, a Chinese philosopher named Chu Shi rewrote
Confucian doctrines within the framework of Taoism.  The entrance
lobby of the main building (commonly called Stalin Tower) of Moscow
State University is decorated with a set of plaques of the world's
greatest scholars, including Socrates, Plato, Aristotle, and Marx.
Chu Shi is also included in this set of wise people.  His version of
Confucianism, commonly called neo-Confucianism, is a strong philosophy
governing all aspects of government and social order, based on Taoist
understanding of the universe.

In 1393, Korea went through a revolution, and a new dynasty emerged
with a new governing philosophy, namely neo-Confucianism.  It was
two hundred years after Chu Shi wrote his books in Fukien province
(south-east) of China.  This new Korean dynasty lasted for more than
500 years until 1910.  It is generally agreed that this longevity
is due to its ideology.  Even though there are many Christian churches
there, Korea can still be regarded as a Confucian-Taoist country.  Its
national flag came from one of the Taoist drawings addressing the
symmetry of our universe.  It could also be said that Korea's recent
progress in democracy and capitalism is largely due to its academic
tradition inherited from this neo-Confucianism.

The story is the same for Japan.  Japan sent many students to China
during the Tang period, and Kyoto was initially copied from Tang's
capital city called {\it Chang An}.  When the Tokugawa family was
in charge of Japan from 1603 to 1870, their governing ideology was
neo-Confucianism formulated by Chu Shi.  While Japanese disagree,
Koreans say that Japan imported neo-Confucianism from Korea by
kidnapping Korean Chu-Shi scholars when Japan's Toyotomi Hideyoshi
invaded Korea in 1593.

Toyotomi Hideyoshi was the first person to unify Japan, but he was
not satisfied with the Japanese islands.  He fancied himself to be
the ruler of the big land, namely China.  In order to crown himself
as the emperor of China, he sent his troops to Korea in order to make
his way to China.  His troops fought in Korea for seven years.  Even
though his troops committed atrocities beyond human imagination to
Koreans, he remains as a very interesting person among Koreans.

Toyotomi is of course remembered by Japanese as a great man.  The
Osaka Castle was originally built by him, but was burnt down by
Tokugawa's troops during the power transition.  The Tokugawa family
then built a bigger castle to honor him.  Since he came from the
lowest social class, there are many jokes about him.  One of them
says that he looks like a human being if monkeys look at him, while
he looks like a monkey if humans look at him.  This definitely is a
Taoist-style joke.  It is also a Kantianist joke.

Taoism is not only a tool for politicians but also provides a basis
for scientific reasoning.  If there is one side, there must be the
opposite side.  Thus, there are two walls in one-dimensional world,
one on one side and one on the other side.  Our three-dimensional
universe is therefore bounded by six walls.  We use this concept of
box normalization in modern physics, especially in kinetic theory of
gases as well in quantization of fields.

Japan's Yukawa Hediki was quite fond of Taoism and studied
systematically the books by Laotse and Chuangtse who were the founding
fathers of Taoism~\cite{tani79}.  Many younger Japanese physicists
used to complain that Yukawa was too philosophical, and did not listen
to him.  It is not unusual for children not to listen to their parents.
While this was going on, I studied Yukawa's papers very carefully.

During the later half of the 19th century, Japanese studied Western
ideology diligently.  Fukuzawa Yukichi was quite interested in the
constitutional monarchy practiced in England.  This is why Japan has
a government similar to that of Britain.  Fukuzawa is also known as
the founder of Keio University, and his portrait is on the Japanese
currency for 10,000 Yens (approximately 80 USD).  During this period,
Japanese scholars translated many books from the West.  One of them
was ``Das Kapital'' written by Karl Marx.  While Chinese disagree,
Japanese say that many Chinese became communists by reading the
Japanese version of Das Kapital.

Likewise, they translated books written by Immanuel Kant.  In the
case of Kant, Japanese not only absorbed his philosophy, but also
extended it to make their own version of Kantianism.  Japanese were
able to do this because they already had their own philosophical
base similar to that of Kantianism, namely Taoist base.  We shall
return to this story in Sec.~\ref{einam}.

During the period from 1910 to 1945, many Koreans went to Japan to
study.  They studied Western ideologies including Marxism and
Kantianism by reading books written in Japanese.  With the same
Taoist base, Koreans were quick to appreciate Kantianism.

Because Japan and Korea share the same philosophical base, it is
very easy to understand and appreciate papers written by traditional
Japanese scholars such as Yukawa.  My publication record will indicate
that I studied Yukawa's papers before becoming seriously interested
in Wignerism.  Indeed, I picked up a signal of possible connection
between Kantianism and Taoism while reading Yukawa's papers carefully,
and this led to my bold venture to ask Wigner whether he was a
Kantianist in 1989, and to my recent venture to Kaliningrad.

I would like to stress that Taoism is not confined to the ancient
Chinese world.  It forms the philosophical base for Sun Tzu's classic
book on military arts~\cite{sonja96}.  Sun Tzu is a very popular
figure among young people in the United States these days.  My maternal
grandfather was fluent in the Chinese classic literature, and he was
particularly fond of Sun Tzu.  He told me many stories from Sun Tzu's
books.  This presumably is how my Taoist background became stronger
than those of other Asian scientists.

Sun Tzu's form of Taoism is practiced frequently in the United States.
Let us look at American football games.  The offensive strategy does
not rely solely on brute force, but is aimed at breaking the harmony
of the defense.  For instance, when the offensive team is near the end
zone, the defense becomes very strong because it covers only a small
area.  Then, it is not uncommon for the offense to place four
wide-receivers instead of two.  This will divide the defense into two
sides while creating a hole in the middle.  Then the quarter-back can
carry the ball to the end zone.  The key word is to destroy the balance
of the defense.

There are two opposite views of everything in Taoism. Does this mean
that we should take only one view and discard the other?  No!  The
truth is somewhere between them.  Finding the harmony between these
two opposite views is the ideal way to live in this world.  We cannot
always live like Shiyu, nor like Idiot, mentioned in Sec~\ref{illus}.
The key to happiness is to find a harmony between the individual and
the society to which he/she belongs.  The key word here seems to be
``harmony,'' and this is what Taoism preaches us.

To Kantianists, however, it is quite natural for the same character
to appear differently in two different environments, again as the case
of Shiyu and Idiot.  The problem is to find the absolute value from
these two different faces.  Does this absolute value exist?  According
to Kant, it exists.  It exists if we find something common to both.
This sounds very much like Taoism.

Finally, let us examine how Taoism was developed in ancient China.
After the last ice age, China was sparsely populated.  On this vast
land, there were many isolated pockets of small tribes.  Then they
came to the banks of China's great rivers, namely the Yellow and
Yangtze Rivers.  There, they developed agriculture and commerce,
using the fertile lands on the river banks and transportation system
provided by the waterways.  Then there came communication problems
among different groups of people with different languages.  They
started to draw pictures to communicate, and sing songs to express
their feelings toward others.  It is well known that Chinese characters
came from drawings.  It is also well known that Chinese, unlike other
languages, has tones.  Their spoken language has tones because of its
musical origin.

In addition, those Chinese had to develop the skills of accommodating
the views of others.  This became Taoism.  Taoism was not dictated by
a single philosopher or group of philosophers.  It became a philosophy
because philosophers attempted to document the way how the people think.
This aspect is strikingly similar to the way in which Kantianism was
developed from the lifestyle of Koenigsbergers.

\section{Einstein and American Physicists}\label{einam}
Einstein came from Germany to Princeton in 1933, and he lived there
until 1955.  He liked America and became a American citizen in 1940.
Yet, Einstein did not communicate well with American physicists.
People are wondering why?

In order to give my answer to this problem, I have to confess that
I have been having problems with Physical Review referees since 1970
when I started asserting my own views in physics.  I still have
problems, but my important papers have already been published in the
appropriate sections of the Physical Review.  For this reason, I can
entertain myself these days whenever I receive hostile reports from
the referees.  Whenever I receive those reports, I assume Einstein
had the same problem with American physicists.

This does not mean that I do not respect American physicists and
their tradition.  I learned about Thomas Edison before Einstein.
Not many American physicists know, but I know that Lee de Forest
found out in 1906 that the grid in a vacuum tube can regulate the
flow of current from the anode to cathode.  De Forest never found
out that this has something to do with the traffic of electrons,
but his discovery was the starting point for electronic industry
which changed the entire world.  De Forest was an excellent American
physicist, but he would have had difficulty in talking to Einstein.

In my opinion, Thomas Edison was the best American philosopher.
Edisonism is not restricted to experimentalists.  Like all American
physicists, I write papers when I do not have ideas.  I obtain new
results while writing papers, and this is true even for the present
paper.  While writing this paper, I got an idea that Taoism developed
in ancient China has the geographical origin similar to that of
Kant's Koenigsberg.

Then, is there a documented version of Edison's philosophy?  Yes, let
us look at the Gospel of Matthew in the New Testament.  Go to Ch.7,
vrs. 7 and 8.  {\it Ask, and it shall be given you; seek ye shall
find; knock, and and it shall opened unto you.  For every one that
asketh receiveth; and he that seeketh findeth; and to him that
knocketh it shall be opened.}  The New Testament is a very important
philosophical document if we insist on philosophy.  The only problem
with the Bible is that it is too easy to read.

Since I have been in the United States since 1954 after my high school
graduation, since I have been teaching American students since 1962,
and since I have an adequate number papers published in the Physical
Review, I am fully qualified to regard myself as a bona-fide American
physicist.  Yet, when I think and write, I am relying on my Eastern
tradition which I explained in Sec.~\ref{tao}.  I become excited when
I find an explanation of fast-moving particles in terms of what
happens to them when they are rest.  I usually get referee comments
that I do not say anything new for particles at rest when I talk about
those at rest.  They are not able to sense the other side of my story,
namely what happens to observers on bicycles.

More recently, I become excited when I find out what happens in optics
in terms of what I know from particle physics.  Thus, to American
particle physicists, I look like an optics man, not one of them.  To
optics people, I am not one of them, because my papers start with what
I learned in particle physics.  I am in a situation similar to Japan's
Toyotomi Hideyoshi mentioned in Sec.~\ref{tao}, who looks like a monkey
to humans while looking like a human being to monkeys.

I was of course aware of this problem and have been looking for a
solution for many years.  I am still continuing my effort along this
direction.  Of course the solution consists largely of understanding
of the cause of the problem.  My tentative conclusion was that I am
a Kantianist while Americans in general are not, but I was eager to
find a convincing evidence to support my assertion.

This is precisely why I went to Kaliningrad last summer even though
the trip was inconvenient if not dangerous.  I visited the Kant
museum twice while staying there.  The museum dedicated one room
for the books written about Kant's philosophy.  There are many books
written in German and many in Russian.  This is not surprising because
Kant wrote his books in German, and the museum is under Russian
management.  What is surprising is that there are many books written
in Japanese.  There are also Japanese scrolls hanging on the wall.
These scrolls contain quotations from Kant's books written in poetic
Japanese.

Since many people in the world write books in English, one would
expect some Kant-related books written in English.  I could not find
any.  I was surprised, but not surprised from my experience with
English-speaking people in the United States.  This is a clear evidence
that, while there is a resonance frequency in the Japanese way of
thinking tuned to Kantianism, there are none in the Anglo-Saxon culture.
I knew this before, but I was able to confirm this in Kaliningrad.
This was one of the happiest moments in my life.  As I said before,
Japan and Korea share the same philosophical base, and I am very proud
to be a Kantianist.

I was happy also because I can now tell why Einstein was not able
to communicate with American physicists.  Einstein was a Kantianist,
while Americans are not.  Americans are Edisonians, and they have their
reason to be proud of themselves.  I have no complaints because I am
also an American physicist in the Edisonian tradition, and the
United States has been very nice to me.

\section*{Concluding Remarks}

In this paper, we attempted to understand Kantian influence on
Einstein in terms of the geography and history of Koenigsberg.  It
is possible to understand the problem without reading Kant's books
if we use the method of physics, namely abstraction from the things
we observe in the real world and then apply the abstraction to
predict what would happen in other parts of the real world.  Using
the same method, we found what happened in ancient China from what
happened in Kant's land of Koenigsberg.

Some of my younger Asian colleagues complain that they are
handicapped to do original research because of the East-West
cultural difference.  I disagree.  This difference could be the
richest source of originality.

I decided to write this article because it is too difficult to read
Kant's books.  I complained to one of my German friends that there
are only two sentences on one page of his books.  My friend disagreed.
He said Kant's one sentence usually covers two pages.  Thus, the best
way to understand his philosophy to write my own articles about him,
in the tradition of Thomas Edison.

\section*{Appendix}

I started doing my research presented in this paper while I was
reading a book review written by by Marie Arana-Ward of the
Washington Post in November of 1994.  She was talking about a book
by Anne Applebaum entitled ``Between East and West, Across the
Borderlands of Europe'' (Pantheon, New York, 1994).  Since this
review is much shorter than the book and explains the geographical
origin of Kantianism, I append her article to this paper with her
permission.

THERE ARE FEW PLACES on earth as politically volatile as the
lands that lie between Russia and the rest of Europe. Both coveted
and despised, this protean corridor of unobstructed terrain has
been a stomping ground for every grandiose politician - East or
West - whose imagination was ever kindled by dreams of conquest.

``For a thousand years, the geography of the borderlands dictated
their destiny,'' writes Spectator deputy editor Anne Applebaum in
her probing portrait of the territory that embraces eastern Germany,
Lithuania.  Belarus, Moldova, Eastern Poland and the Ukraine. ``Five
centuries ago an army on horseback could march from a castle on the
Baltic to a fort on the Black Sea without meeting a physical obstacle
greater than a fast-running river or a wide forest. Even now, a spy
running east from Warsaw to Kiev would find nothing natural to
obstruct him.''

And so for centuries the people of these frontiers endured carnage
and plunder, slipping in and out of identities faster than a
cartographer could record the changes: Mongols invaded, and then
Turks, Swedes, Muscovites, Moldovan princes, Cossacks, Teutonic Knights,
Polish kings, German emperors, Nazis, Soviet hordes - each raid more
catastrophic than the last. The Swedes destroyed the cities, the
Cossacks set fire to the villages, the Teutonic Knights brought on a
holocaust, wiping out all trace of indigenous Prussians. "But most of
the time," writes Applebaum, "the Polonizations and Prussifications
and Russifications came to nothing. The borderlands were simply too
wide and too empty, it was too difficult for any invading nation
to maintain permanent rule."

Because of this failure to bring about long-term change there were,
until recently, no nation-states as we know them today.  ``For a
thousand years the people of the borderlands spoke their dialects
and worshiped their gods, while the waves of invaders washed over
them, mingled, receded, and washed over them again.''  Today a traveler
can encounter a native Pole, a person raised in the Soviet Union, a
citizen of the new Belarus - and they in fact may all be one man, an
individual who has never set foot outside his father's village.

A borderland peasant asked his nationality in the 18th century
probably would have replied {\it tutejszj - ``a person from here.''}
That  sense of the existential still persists.  A scene in Gunter
Grass's {\it The Tin Drum } captures the mindset: It is 1945, and
as Soviet soldiers pour into the borderlands, terrorizing the locals,
the protagonist's grandmother refuses to flee east or west, ''I am
not German enough!'' she cries. ``And I am not Polish enough, either!''
She belongs to her potato  field; it is the only allegiance she
considers worthwhile.

After the war, it was Stalin's plan to have the borderlands ``disappear
into Soviet Russia: Call it ethnic cleansing, to use a phrase coined
later in another context, or call it cultural genocide. Either way,
it was very successful.''  The region was transformed beyond recognition.
Whole nations slid beyond memory, and we in the West hardly took notice,
Kiev became a Russian city, Lithuania a Russian province, and the
colorful, variegated cultures of the borderlands were relegated to
the dusty shelves of emigre bookshops.

Applebaum's {\it Between East and West} is a heroic attempt to bring
the region back into our collective consciousness. Armed with 35 maps
and a ``forensic passion,''
she leads us into this forgotten land, holding a close mirror to its
villains and heroes and letting us see it warts and all.

She begins, fittingly, in Kaliningrad - Koenigsberg - a district once
famed as the City of Enlightenment - Kant's city - before it was purged
of Prussians and reravaged centuries later by Soviet troops.

``Sprechen Sie Deutsch?'' someone asks her on the streets
of the now hideous city. The man is a Belarusian from Pinsk, a slave
laborer in a German prison factory during the war. It is the first of
a multitude of encounters that will lead Applebaum - and us - to a
clearer understanding of who these people are and how history has
transformed their lives.

In Lithuania and the Polish {\it kresy - the outlying, disputed
hinterland}, Applebaum encounters the core of hatred that has set
neighbor against neighbor for 50 years.  Before the war, some Jews,
encouraged by Soviet propaganda, had collaborated with the communists;
some Lithuanians, encouraged by the Nazi propaganda, had helped send
Jews to concentration camps. ``Afterward,'' writes Applebaum,
''no one remembered that the Red Army had also murdered Jews, or
hat the Nazis had also murdered Lithuanians. "The outcome had been
too dire to parse history that finely: One in ten Lithuanians was
either dead or deported.  Several million Poles were forcibly removed
from their homes in Lithuania, Belarus and the Ukraine, and sent West.
"When it was over, the mixed multiethnic {\it kresy } had disappeared
forever. Most of the Poles were gone from
the region, most of the Jews were dead.'' But
the bitterness remains.

Applebaum negotiates the region intrepidly, suffering
the hardscrabble existence of a traveler in these parts,
looking up improbable witnesses, hitch-hiking with drink-sodden
peasants, arguing history with strangers on the street.
Her insights are sharp, her sympathies far-ranging.
Always there is an unblinking eye on her subjects'
life stories, a finger on history, and a well-tuned ear for
subtle ironies and unexpected poetry.  Whether negotiating
with a slick Mafia hotel manager in L'vov or bedding down with
a pestiferous anti-Semitic harridan in Nowogrodek, Applebaum
reveals an intelligence and sensibility that are rare in this
brand of quick-sweep expeditionary journalism.

But the book is not free of flaws. As we progress from
Kamenets Podolsky to Kishinev to Odessa, we sense a
progressive impatience in our host.  When she wraps up her
trip and boards a boat for Istanbul and "the West" with a
distinct sense of relief, we cannot help but recall her
initial statement of purpose: that what she had set out to find
was "proof that difference and variety can outlast an imposed
homogeneity; testimony, in fact, that people can survive
any attempt to uproot them."   Applebaum's rush to be
done with her book ultimately leaves her reader hanging on that
question. We understand that history has made the people of the
borderlands at once indomitable and chameleonesque, but that is
our conclusion; Applebaum never tells us hers.

That said, {\it Between East and West} is an indispensable guide
to a little-known region that may prove as decisive in our
future as it surely has been in our past.  As Churchill wrote
when the various nations of the borderlands first proclaimed
their independence, ``When the war of the giants has ended, the
war of the pygmies begins.''  We would do well to know the territory

\end{document}